\documentclass[floatfix,a4paper,prl,twocolumn,superscriptaddress, showpacs,amsmath,amssymb,footinbib]{revtex4-1}
\pdfoutput=1

\usepackage{natbib}
\usepackage[english]{babel}
\usepackage{letltxmacro}
\LetLtxMacro{\ORIGselectlanguage}{\selectlanguage}
\makeatletter
\DeclareRobustCommand{\selectlanguage}[1]{%
  \@ifundefined{alias@\string#1}
    {\ORIGselectlanguage{#1}}
    {\begingroup\edef\x{\endgroup
       \noexpand\ORIGselectlanguage{\@nameuse{alias@#1}}}\x}%
}
\newcommand{\definelanguagealias}[2]{%
  \@namedef{alias@#1}{#2}%
}
\makeatother
\definelanguagealias{en}{english}
\definelanguagealias{English}{english}

\usepackage{graphicx}
\usepackage[usenames,dvipsnames]{color}

\usepackage[colorlinks=true,allcolors=blue]{hyperref}

\usepackage{bbm}

\newcommand{\tr}[2][]{\ensuremath{\text{tr}_{#1}\left[ #2\right]}}

\newcommand{\mean}[1]{\left\langle #1 \right\rangle}	

\newcommand{\abs}[1]{\ensuremath{\left\vert #1 \right \vert}} 
\newcommand{\bra}[1]{\langle #1 |}
\newcommand{\ket}[1]{| #1 \rangle}

\newcommand{\matrixel}[3]{\ensuremath{\left\langle #1 \middle \vert #2 \middle \vert #3\right\rangle}}

\newcommand{\tot}[1]{\mathrm{d}\hspace{-0pt}#1\;}

\def\clap#1{\hbox to 0pt{\hss#1\hss}}

\begin{document}

\title{Detection and characterization of Many-Body Localization in Central Spin Models}

\author{Daniel Hetterich}
\affiliation{Institute for Theoretical Physics,  University of W\"urzburg, 97074 W\"urzburg, Germany}

\author{Norman Y.~Yao}
\affiliation{Department of Physics, University of California, Berkeley, California 94720, USA}

\author{Maksym Serbyn}
\affiliation{Institute of Science and Technology, 3400 Klosterneuburg, Austria}

\author{Frank Pollmann}
\affiliation{Department of Physics, Technical University Munich, 85748 Garching, Germany}

\author{Bj\"orn Trauzettel}
\affiliation{Institute for Theoretical Physics,  University of W\"urzburg, 97074 W\"urzburg, Germany}

\date{\today}

\begin{abstract}
We analyze a disordered central spin model, where a central spin interacts equally with each spin in a periodic one dimensional random-field Heisenberg chain. If the Heisenberg chain is initially in the many-body localized (MBL) phase, we find that the coupling to the central spin suffices to delocalize the chain for a substantial range of coupling strengths. We calculate the phase diagram of the model and identify the phase boundary between the MBL and ergodic phase.  Within the localized phase, the central spin significantly enhances the rate of the logarithmic entanglement growth and its saturation value. We attribute the increase in entanglement entropy to a non-extensive enhancement of magnetization  fluctuations induced by the central spin.     Finally, we demonstrate that correlation functions of the central spin can be utilized to distinguish between MBL and ergodic phases of the 1D chain. Hence, we propose the use of a central spin as a possible experimental probe to identify the MBL phase. 
\end{abstract}

\maketitle
\emph{Introduction.---}Many-body localization (MBL) is the interacting analog of Anderson localization \cite{Anderson1958,Basko2006a}.  As localized systems are perfect insulators, they violate the eigenstate thermalization hypothesis (ETH) \cite{Deutsch1991,Srednicki1994}.  This violation implies that expectation values of physical observables with respect to eigenstates may not be described by thermodynamic ensembles anymore. Hence, the characteristic repulsion between energy levels of typical thermalizing systems~\cite{Wigner1955} is absent in the MBL phase. The absence of level repulsion and the intrinsic memory about the initial state in the MBL phase may be understood via an emergence of local integrals of motion~\cite{Serbyn2013a,Huse2014}. ETH can also be violated in systems that do not experience MBL, such as integrable systems \cite{Berry1977}. However, in contrast to integrable systems, the MBL phase is stable to weak but finite local perturbations~(see \cite{Abanin2018} for a recent review). Moreover, the signatures of MBL can be observed in presence of weak coupling to heat baths~\cite{Levi2016} and particle loss~\cite{VanNieuwenburg2017}. However, the robustness of MBL exposed to long-range interactions is still an open question. While it has been proposed that MBL  exists in systems with interactions  that decay with distance as a power law \cite{Burin2006,Yao2014}, in a recent work it is argued that MBL could be present in systems with non-decaying interactions~\cite{Nandkishore2017}.

In this paper, we study the behavior of the MBL transition in the presence of a central spin that equally couples to all other spins in the model \cite{Gaudin1976}. Our model therefore obtains a very particular type of long-range interaction, in which each spin is effectively coupled to all other spins via the central spin. 
 These models are experimentally relevant for spin qubits based on electrons captured in quantum dots \cite{Loss1998}. In such systems, a qubit plays the role of the central spin that experiences decoherence due to the environmental bath spins \cite{Uhrig2007,Lee2008}. The central spin interacts with the bath of nuclear spins  via hyperfine interaction~\cite{Coish2004,Fischer2009} which was experimentally investigated in different host materials~\cite{Hanson2007}. Similarly, nitrogen vacancies in diamond represent central spins whose main source of decoherence are electron spins of surrounding nitrogen impurities \cite{Jelezko2004,Hanson2006}.

The main result of this work is that a central spin can be employed in order to detect localization of its environment. To this end, we first study the impact of the central spin on the well-known MBL transition of the Heisenberg chain~\cite{Pal2010,Luitz2015}. We find an analytic expression for the critical disorder at which the transition from MBL to the ergodic phase appears. The central spin establishes a non-local coupling that enhances  the rate of the logarithmic growth of the half-chain entanglement entropy and its saturation value. We observe that this enhancement has the same form as the non-extensive increase in magnetization fluctuations that we find, which suggests a relation between these two effects.
The latter effect was analytically analyzed  in a fermionic non-interacting central site model (NCSM) \cite{Hetterich2017}.   Finally, we propose a novel detection scheme for MBL based on the autocorrelation function of the central spin. We show that the behavior of the autocorrelation function at large frequencies provides information about the state of the environment of the central spin. Thus, it can be exploited as a MBL detector.

\emph{Model.---}We extend the random field Heisenberg chain showing a MBL transition \cite{Pal2010,Luitz2015} by coupling all sites to the central spin:
\begin{equation} \label{eq:ham}
H = J\sum_{i=1}^{K} \vec{I}_i \cdot \vec{I}_{i+1} + \sum_{i=1}^K B_i I_i^z + \frac{A}{K}\sum_{i=1}^K \vec{S} \cdot \vec{I}_i,
\end{equation}
where $\vec{S}=\frac{1}{2}(\sigma_x, \sigma_y,\sigma_z)^T$ is the central spin that equally couples to the $K$ spins $\vec{I}$ of the Heisenberg chain with periodic boundary conditions.  The random fields $B_i$ are uniformly distributed  $B_i \in [-W,W]$, where $W$ sets the disorder strength, and we set $J=1$ in the following. 

For $J=0$, our model becomes similar to a previously studied system~\cite{Ponte2017}, where the l-bit \cite{Serbyn2013a,Huse2014} representation of MBL was employed  to study the influence of a central spin on its MBL environment. The authors of Ref.~\cite{Ponte2017} demonstrated that the l-bits remain localized when their coupling strength to the central spin is rescaled with the inverse system size. Hence, in  Eq.~(\ref{eq:ham}) we choose the coupling of the central spin to the physical spin degree of freedoms to be $A/K$. Such scaling ensures that the spectral bandwidth of the coupling term is independent of system size. Then, the spatially non-local coupling term to the central spin can be considered as being local in energy space. Moreover, a coupling rescaled in this way is experimentally relevant in certain quantum dot models \cite{Loss1998, Fischer2009}, for which we propose below a concrete way to detect MBL. While the relaxation features of similar central spin models have previously been studied \cite{Hetterich2015, Reimann2016}, we focus on the MBL signatures of central spin models in this paper.

\begin{figure}[t]
\centering
\includegraphics[width= 0.99\linewidth]{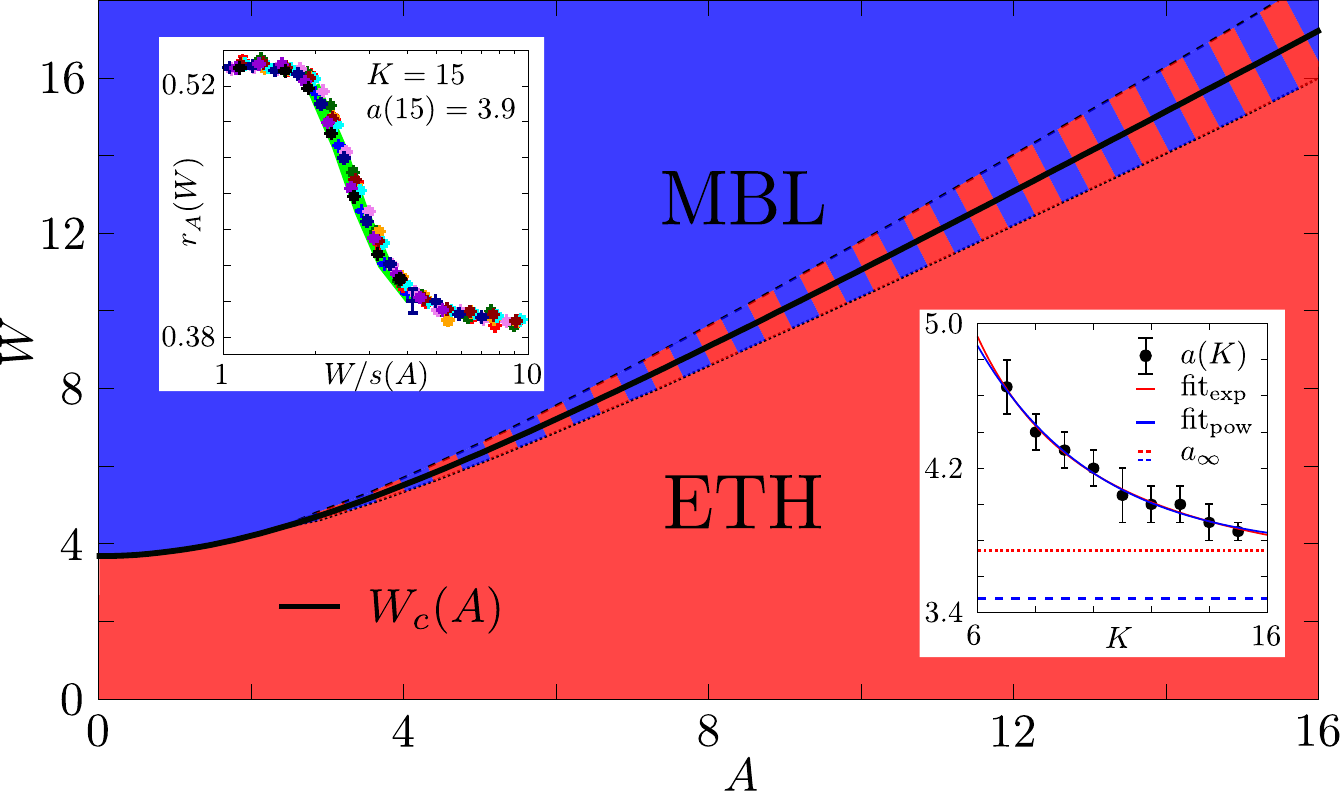}
\caption{Phase diagram of the central spin model. The critical disorder strength $W_c(A)=s(A) \cdot W_c^\text{Heis}  $ (solid line), at which the eigenvalues in the center of the band  transition from a Poisson distribution towards the GOE ensemble, grows with coupling strength $A$ to the central spin. The stripes represent the uncertainty of the parameter $a_\infty$, which arises from the extrapolation extrapolation of $a(K)$ to the thermodynamic limit, see right inset. We find $a_\infty = 3.55\pm 0.25$, where the uncertainty arises by comparing  power-law or exponential fitting functions. Each value $a(K)$ is obtained by a scaling analysis as illustrated in the left inset, where the disorder strength is rescaled by $s(A)$ for all simulated values of $A$.}
\label{fig:phasediagram}
\end{figure}

\emph{Phase diagram of the central spin model.---}An efficient way to distinguish ergodic and localized phases is to exploit their different eigenvalue statistics. While eigenvalues repel each other in the ergodic phase, leading to a Gaussian Orthogonal Ensemble (GOE) of levels, eigenvalues are simply Poisson-distributed (POI) in localized phases. Both phases lead then to  different distributions of gaps $g_i=E_{i+1}-E_i$ of adjacent energies. A commonly used indicator of level statistics is the ratio of adjacent energy gaps,  $r_A(W) = \mean{\min(g_i,g_{i+1}) / \max(g_i,g_{i+1})}_i$ \cite{Oganesyan2007}, which takes values between $\approx 0.53$ (GOE) and $0.38$ (POI). The average runs over disorder ensembles and eigenvalues in the center of the spectrum. Since the bandwidth of the terms responsible for coupling to the central spin is limited, their effect on the levels $E_i$ of the Heisenberg chain crucially depends on the position in the spectrum. We focus on levels in the center of the band, where the density of states is largest and one expects the onset of delocalization. 
   
In the absence of the central spin, the model is known to show a MBL transition at $W_c(A=0)=W_c^\text{Heis}\approx 3.7$~\cite{Pal2010,Luitz2015}. Upon increasing $A$ we find that $r_A(W)$ is well approximated by $r_A(W) =  r_0(W/s(A)) $, where $r_0(W)$ is the value of the indicator $r$ for the pure random field Heisenberg chain. The rescaling function 
\begin{equation}\label{eq:shift}
s(A)=\sqrt{1+(A/a)^2},
\end{equation}
depends on a single parameter $a$ that changes with system size but does not depend on the disorder strength~\cite{Note2}.  The form of rescaling function in Eq.~\eqref{eq:shift} is motivated by limits found in previous works. For small values of coupling $A$ Eq.~\eqref{eq:shift} recovers the result of the random field Heisenberg chain with a second order corrections similar to the case of the NCSM~\cite{Hetterich2017}. On the other hand, for $A\gg 1$ we recover $W_c(A) \approx A$, consistent with predictions of Ref.~\cite{Ponte2017}.

The quality of the rescaling collapse is shown in the left inset of Fig.~\ref{fig:phasediagram}, where the results for many different coupling constants $A$ are mapped onto the known result of the random field Heisenberg chain.
The asymptotic value of the free parameter as $K\to \infty$ is determined to be $a= 3.55\pm 0.25$. The finite size scaling analysis is shown in the right inset of Fig.~\ref{fig:phasediagram}. Finally, Fig.~\ref{fig:phasediagram} illustrates the resulting critical disorder strength
\begin{equation}
W_c(A) = W_c^\text{Heis} s(A),
\end{equation}
which separates the localized from the ergodic phase. We want to emphasize that, for a given disorder strength $W>W_c^\text{Heis}$, the central spin needs to couple sufficiently strong in order to delocalize eigenstates in the center of the band. This result is a clear many-body effect, because, for the NCSM, we have found an energy window of size $\sim A^2/K$ consisting of repelling eigenvalues at any $A>0$~\cite{Hetterich2017}.

\begin{figure}
\centering
\includegraphics[width= 0.99\linewidth]{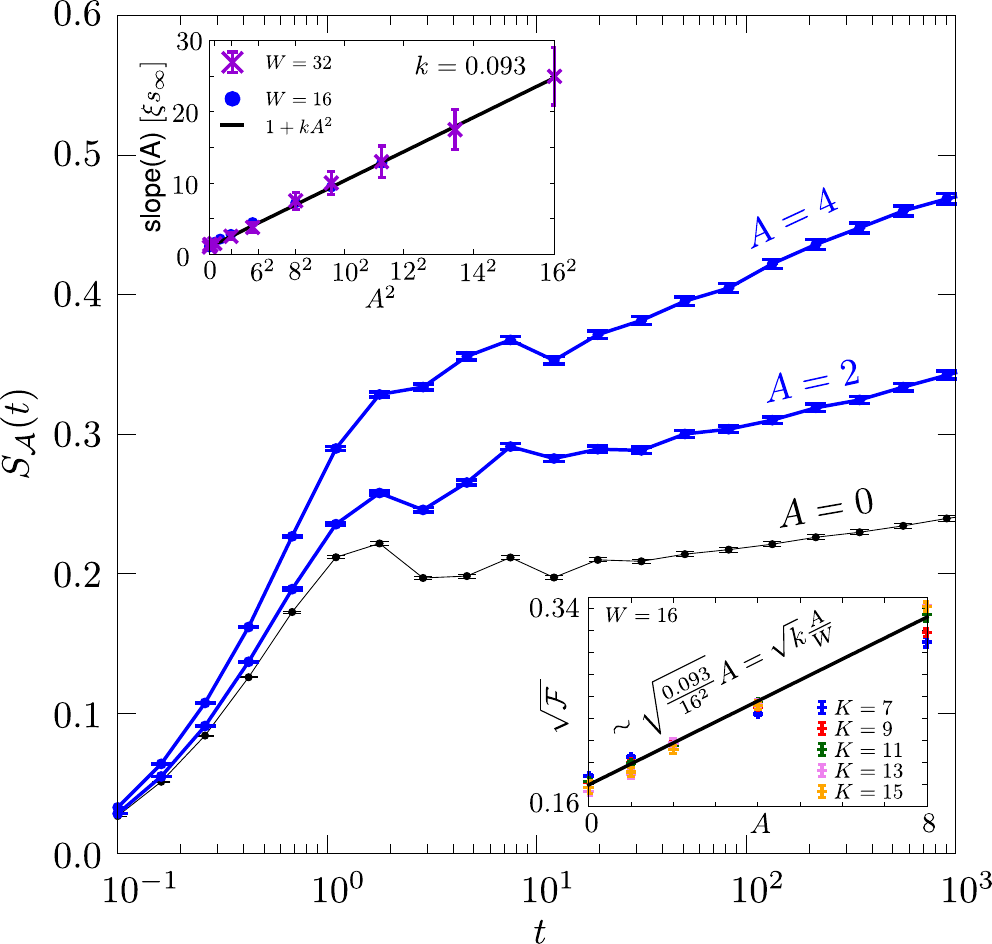}
\caption{ Growth of entanglement entropy $S_\mathcal{A}(t)$ of the N\'eel state for different coupling constants to the central spin. In the localized phase, we find that the slope of the logarithmic entanglement growth increases quadratically with $A$ (see upper left inset), which motivates Eq.~(\ref{eq:entanglement}). For the fit parameter $k$ we  find $k\approx 0.093(5)$.
The bottom right inset shows the fluctuation $\mathcal{F}$ (see text) of eigenstates inside the localized phase. We find a non-extensive behavior of $\mathcal{F}\sim k A^2/W^2$, which, as $\xi\sim 1/W^2$, traces the enhancement of $S_\mathcal{A}(t)$ back to magnetization exchange between bipartitions.
The data is generated for $W=16$ using $14$ spins.
}
\label{fig:ent}
\end{figure}

\emph{Logarithmic growth of entanglement entropy.---} The logarithmic growth of entanglement entropy is employed as a signature of the interacting localized phase with  local Hamiltonians~\cite{Znidaric2008,Bardarson2012}. At the same time, the non-local NCSM also displays logarithmic growth of entanglement entropy despite the absence of interactions~\cite{Hetterich2017}. Therefore, it is instructive to study the dynamics of entanglement entropy in the interacting central spin model. Starting with the N\'eel state $\ket{\psi(t=0)}=$ $\ket{\uparrow\downarrow \uparrow\ldots}$, we compute the reduced density matrix $\rho_\mathcal{A} = \tr[\mathcal{B}]{\ket{\psi(t)}\bra{\psi(t)}}$, where we trace out $K/2$ contiguous spins. As the entanglement entropy is $S_\mathcal{A} = S_\mathcal{B} = -\tr{\rho_\mathcal{A} \ln \rho_\mathcal{A}}$, the result is independent of which bipartition contains the central spin.  For coupling strength $A=0$, we recover the case of a periodic Heisenberg chain.   Here, $S^\text{Heis}_\mathcal{A}(t) \sim \xi s_\infty \ln t$ grows logarithmically in time, where $\xi$ is the localization length of the model in the absence of interactions  and $s_\infty$ the contribution to the saturation value of $S_\mathcal{A}(t)$ per spin \cite{Serbyn2013}. Figure~\ref{fig:ent} shows that non-zero coupling  to the central spin increases the rate of the entanglement growth as:
\begin{equation}\label{eq:entanglement}
S_\mathcal{A} \sim  \xi s_\infty \left( 1 +  k A^2 \right)\ln t,
\end{equation}
where $k$ is a constant that is independent of $W$ and $A$. Note that the slope of the logarithmic entanglement growth may be completely dominated  by the central spin (left inset of Fig.~\ref{fig:ent}). 
Equation~\eqref{eq:entanglement} can be rewritten as 
$S_\mathcal{A} = \tilde{\xi} \tilde{s}_\infty \ln t$, 
where $\tilde{\xi}$ and $\tilde{s}_\infty$ are the effective correlation length and the saturation entropy density in the presence of the central spin.

The enhancement of the logarithmic entanglement growth originates from an increase in both $\tilde \xi$ and $\tilde{s}_\infty$ compared to $\xi$ and $s_\infty$, as we discuss in the supplementary material \cite{Note2}. The functional form of the enhancement coincides with the enhancement of fluctuations  of magnetization $\mathcal{F}$ between the considered bipartitions. More specifically, for 
 $\mathcal{F} = \mean{{J^z_{\mathcal{A}}}^2}-\mean{J^z_{\mathcal{A}}}^2$ with the total spin $J^z_\mathcal{A}=\sum_{i\in\mathcal{A}}I_z^i$ inside a bipartition $\mathcal{A}$ for eigenstates in the center of the spectrum, we find the same dependency: $\mathcal{F} \sim k A^2/W^2$ (see right inset of Fig.~\ref{fig:ent}).  We emphasize that $\mathcal{F}$ is not extensive in the localized phase \cite{Note2}, such that the total amount of magnetization `transmitted' through the central spin remains constant if the system size is increased. This critical behavior is necessary for simultaneously maintaining both a constant magnetization exchange and localization at $K\to\infty$. It is a result of the rescaling $A/K$ of the coupling term in Eq.~\eqref{eq:ham}. Notably, we have found the same scaling for the logarithmic transport in the NCSM using second order perturbation theory in $A$. \footnote{In the NCSM, we have chosen the scaling $A/\sqrt{K}$ and derived the motion of a single fermion, where the relevant process was derived in 2nd order perturbation theory $\sim A^2/K$. In this work, we have $K/2$ particles (N\'eel state) instead, which is the reason why a different scaling of the coupling strength, i.e. $A/K$, yields similar results.} While the similar functional dependence suggests that fluctuations of magnetization are responsible for the enhanced growth of entanglement entropy, analytical understanding of the increase in $\tilde \xi$ and $\tilde{s}_\infty$ remains an interesting open question.

We conclude that, at sufficient disorder strength, the central spin model is many-body localized in terms of  thermodynamical and quantum statistical perspectives. Information, witnessed by entanglement entropy, spreads at most logarithmic in time. Eigenvalues are Poisson-distributed and the corresponding eigenvectors have an area-law entanglement entropy~\footnote{See the supporting online material for more details.}. The system fails to self-thermalize and preserves information about the initial state.

\begin{figure}
\centering
\includegraphics[width= 0.99\linewidth]{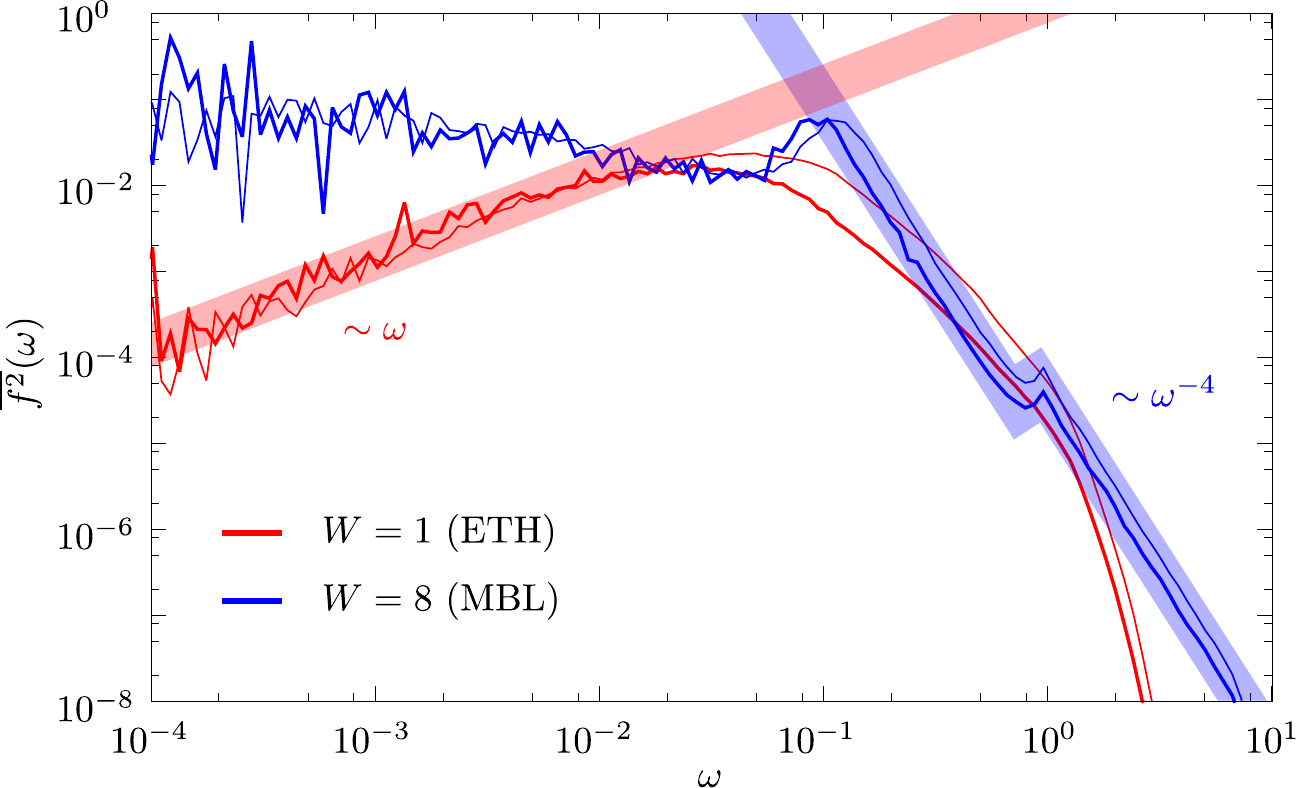}
\caption{Fourier spectrum of the autocorrelation function. At small frequencies $\omega \ll A$, the central spin can detect the ETH phase (red) by a linear decay of $\overline{f^2}(\omega)$, which originates from level repulsion. 
In the MBL phase, $\overline{f^2}(\omega)$ shows a significant power-law decay over many orders of magnitude. The exponent $-4$ is independent of system size, disorder, and coupling strength. The two peaks at $\omega\sim 10^{-1}$ and $\omega\sim 1$ correspond to the local interactions $A$ and $A/K$, which are revealed in the dynamics of the central spin.  Data is generated for 10 (thin lines) and 12 (thick lines) spins at $A=1$. The colored areas are guides for the eye and indicate the power-law behaviors.}
\label{fig:eft}
\end{figure}

\emph{Detecting MBL with the central spin.--} After we have demonstrated that there exist systems in which the insertion of a central spin does not destroy the MBL phase, we explain how the central spin can be used as an ideal (non-demolition) detector of MBL. In particular, we assume that the measurable quantity is a spin component of the central spin, e.g. $S_z(t)=\matrixel{\psi(t)}{S_z}{\psi(t)}$. We investigate its autocorrelation function
\begin{align}\label{Eq:Ct}
C(t) &= \int_{-\infty}^\infty \tot{\tau} S_z(t+\tau) S_z(\tau)\\
&= \sum_{nm} \abs{\rho_{nm}^E}^2 \abs{(S_z^E)_{nm}}^2 e^{i(E_n-E_m)t}, \nonumber
\end{align}
where $S^E_z$ and $\rho^E$ are the observable and the initial density matrix in the energy space of eigenstates with energies $E_n$.
The Fourier transform of Eq.~(\ref{Eq:Ct}) yields
\begin{align}
f^2(\omega) &= \frac{1}{2\pi} \int_{-\infty}^{\infty} e^{-i\omega t} C(t)\\
&= \sum_{nm} \abs{\rho_{nm}^E}^2 \abs{(S_z^E)_{nm}}^2 \delta[\omega - (E_n-E_m)].\nonumber
\end{align}
Note that $f^2(\omega)$ is frequently studied in the context of the ETH  \cite{Srednicki1999} and is thus a natural candiate for helping to identify localization~\cite{DAlessio2016}. Evidently, $\rho_z^E$ and $S_z^E$ can only contribute to $f^2(\omega)$ if there exists two energies with $\omega = E_i-E_j$ exists. The energies $E_i$ and $E_j$ are not limited to be adjacent energy levels, but yet, the behavior of $f^2(\omega)$ for $\omega\ll \mean{\delta_i}$ is dominated by the statistics of level spacings $\delta_i = E_{i+1} - E_i$. In particular, in the ergodic phase, where eigenvalues repel each other, the probability to find a small level spacing behaves as $p(\omega) \tot{\omega} = ({\pi}/{2}) \omega e^{-{\pi} \omega^2/4} \tot{\omega}\propto \omega \tot{\omega.}$ Therefore, in contrast to the localized phase, we expect that  $f^2(\omega)$ is linearly suppressed in the ergodic phase. The dynamics of the central spin is hence influenced by the level statistics of the surrounding spins. We illustrate this feature in Fig.~\ref{fig:eft}, where we present the disorder average of the smoothed discrete function
\begin{equation}
\overline{f^2}(\omega_i) = \frac{1}{\Delta(\omega_i)}\int_{\omega_i}^{\omega_i+\Delta(\omega_i)} \tot{\omega}f^2(\omega).
\end{equation}
We indeed find $\overline{f^2}(\omega)\sim \omega$ in the extended phase at small frequencies $\omega \ll A/K$.

Above we have demonstrated that the presence or absence of level repulsion manifests in a qualitatively different behavior of $\overline{f^2}(\omega)$ at frequencies of the order of the level spacing, hence allowing to distinguish between MBL and ergodic phases. In addition, we also observe a qualitatively different behavior of the autocorrelation function at larger frequencies. In the MBL phase, we find clear peaks of $\overline{f^2}(\omega)$ at $\omega=1$ and $\omega=A/K$, corresponding to the coupling strength between neighbored spins of the Heisenberg chain and their coupling strength to the central site, respectively. In that case, the dynamics of the central spin is strongly affected by local interactions, in contrast to the extended phase where we do not see any pronounced features. It should be noted that most weight of $\overline{f^2}(\omega)$ is concentrated in the vicinity of $\omega = A/K$ in the localized phase (this is masked by the logarithmic scale in Fig.~\ref{fig:eft}).

The last and most significant feature is the power-law decay of $\overline{f^2}(\omega)$ in the localized phase for $\omega>A/K$, which ranges (even in our rather small system of 14 spins) over 7 orders of magnitude. A power-law dependence of a  related quantity to $f^2(\omega)$ has recently been studied in terms of localization in Ref.~\cite{Serbyn2017}. We find that the exponent of the power-law is independent of system size (see Fig.~\ref{fig:eft}), disorder strength, and also independent of the coupling strength to the central spin \cite{Note2}. Further, for  different distributions of random numbers, such as normal and lognormal distributions, we have observed the same exponent $p=-4$, which therefore seems to be a generic exponent of this model and a novel indicator of MBL.  

From the experimental side, one possible realization of our model is afforded by nitrogen vacancy (NV) centers in diamond \cite{doherty2013nitrogen, Schirhagl2014}. We envision working with high nitrogen density Type Ib samples, where the dominant defects are spin-$1/2$ P1 centers (nitrogen impurities). In this case, the NV center then plays the role of an optically addressable central spin while the P1 centers play the role of the bath spins. By working at a magnetic field near $B \sim 510\,\text{G}$, the NV and the P1 defects become resonant and dipolar couplings mediate strong interactions between them~\cite{Hall2015}. We note that in this setup, disorder occurs also in the strength of these dipolar interactions, which scale as $1/r^3$. Finally, one should be able to directly measure the central NV's frequency dependent spin-spin autocorrelation function. This can be done via spin-echo like pulse sequences in the range $\omega \sim 10^{-1}\,J$ to $10^2\,J$~\cite{Schirhagl2014}, which can be used to diagnose the presence of a MBL phase.

\emph{Conclusion.---}We have studied dynamical and statistical properties of a  central spin variant of the  Heisenberg model. Using an equal coupling strength $A/K$ to all spins, where $K$ is the length of the Heisenberg chain, the system shows, depending on the disorder strength, either a MBL or ergodic phase. We have identified an analytical function $W_c(A)$ for the critical disorder strength at which the phase transition occurs. In the localized phase, $W> W_c(A)$, we have observed an enhanced logarithmic spreading of entanglement entropy, which induced by a non-extensive exchange of magnetization.  We have proposed to employ the central spin as a detector to distinguish between MBL and  ergodic phase by means of autocorrelation functions.

We would like to thank Fernando Dom{\'i}nguez, David Luitz, Joel Moore, Tommy Schuster, and Niccol\`o Traverso Ziani for insightful discussions and Gregory Meyer for introducing us to his powerful python interface ``dynamite''. Financial support has been provided by the Deutsche Forschungsgemeinschaft  (DFG)
via Grant No. TR950/8-1, SFB 1170 ToCoTronics and the ENB  Graduate  School  on  Topological  Insulators. FP acknowledges the support of the DFG Research Unit FOR 1807 through grants no. PO 1370/2- 1, TRR80, the Nanosystems Initiative Munich (NIM) by the German Excellence Initiative, and the European Research Council (ERC) under the European Union's Horizon 2020 research and innovation programme (grant agreement no. 771537). 
NYY acknowledges support from the NSF (PHY-1654740), the ARO STIR program and a Google research award.

\clearpage

\section{Supporting online material}

\subsection{Numerical method}

We make use of the fact that the Hamiltonian of the central spin model
\begin{equation} \label{someq:ham}
H = \sum_{i=1}^{K} \vec{I}_i \vec{I}_{i+1} + \sum_{i=1}^K B_i I_i^z + \frac{A}{K}\sum_{i=1}^K \vec{S} \vec{I}_i
\end{equation}
commutes with the total spin $z$ operator $J_z = \sum_i I_i^z + S^z$. Hence, we work in the largest subspace with constant $J_z$. We then construct a matrix representation of $H$ within this subspace. We draw the values $B_i$ randomly from the range $[-W,W]$ with a flat probability distribution.
For the logarithmic entanglement growth and the autocorrelation function $f^2(\omega)$, where we need all eigenvalues, we make use of exact diagonalization techniques offered by the python package 'numpy'. For the level statistics in the center of the band, we use a shift-invert method provided by the python wrapper 'dynamite'  of the scalable linear algebra packages PETSc and SLEPc.

\subsection{Level statistics and phase diagram}
We determine around 50 eigenvalues $E_i^n$ in the center of the band. The actual amount may differ in a given disorder ensemble $n$ as the used shift-invert method may return more eigenvalues if more eigenvalues happen to converge. We then evaluate \begin{equation}
r_A(W) =  \mean{\frac{\min(E_{i+1}^n - E_{i}^n, E_{i+2}^n-E_{i+1}^n)}{\max(E_{i+1}^n - E_{i}^n, E_{i+2}^n-E_{i+1}^n)}}_{i,n},
\end{equation}
where we have averaged over all computed eigenvalues and disorder ensembles of each  parameter set $(W,A,K)$. Note that we do not explicitly indicate the $K$ dependency if not necessary. 

\begin{figure}
\centering
\includegraphics[width=0.98\linewidth]{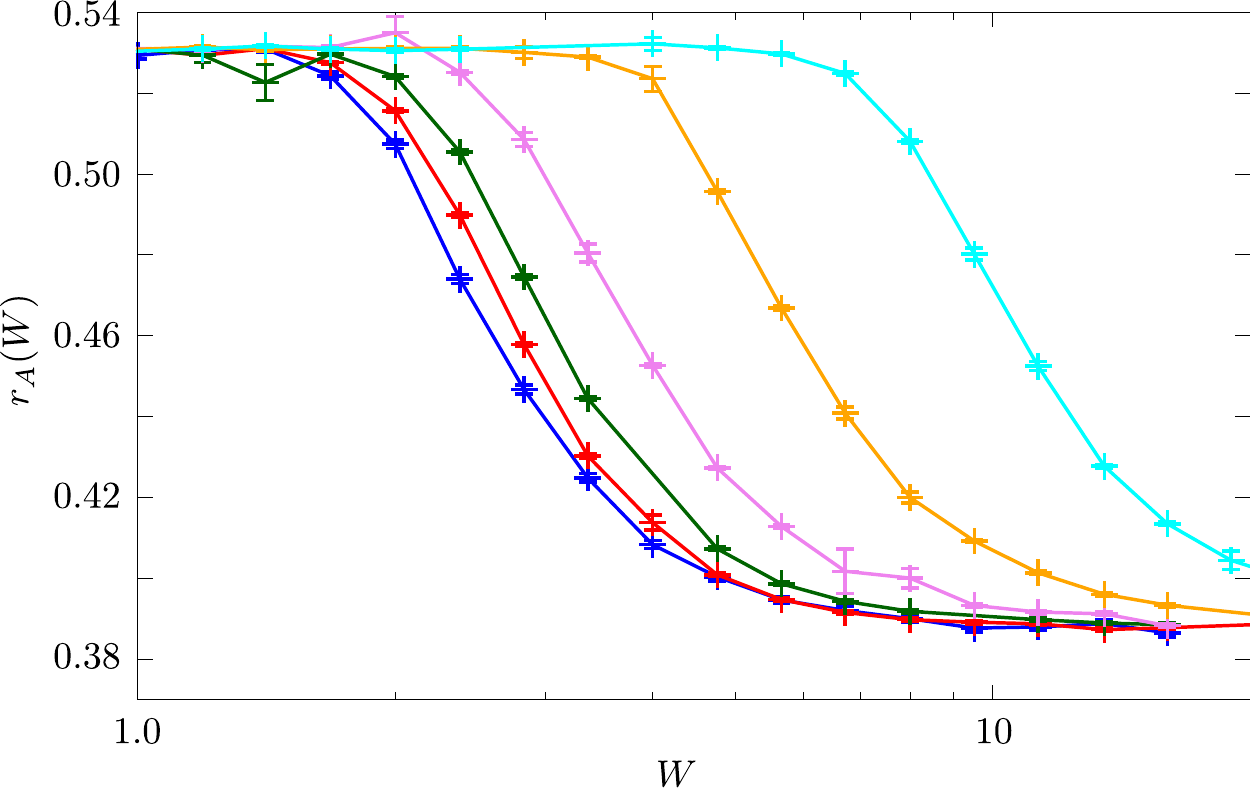}
\caption{Eigenvalue statistics at $K=13$ for $A\in\{0.0,1.0,2.0,4.0,8.0,16.0\}$ (left to right). The functions $r_A(W)$ are shifted towards higher disorder values compared to the random field Heisenberg chain $A=0$ (blue).}
\label{fig:SOM_data_collapse_unshifted}
\end{figure}

Comparing the functions $r_A(W)$ at a given system size $K$ for different values of $A$, we can investigate the impact of the central spin upon the random field Heisenberg chain. For all studied values, higher values of $A$ simply 'maps' the level statistics to larger disorder strengths $W$. This can be seen in Fig.~\ref{fig:SOM_data_collapse_unshifted} and motivates the shifting function $s(A)$, where
\begin{equation}
r_A(W) = r_0(W / s(A))
\end{equation}
at given system size $K$, where $r_0(W)$ is the level statistic of the bare random field Heisenberg chain. The simplest form of $s(A)$ that fulfills the limits described in the main text is $s(A) = \sqrt{1+(A/a)^2}$, where $a=a(K)$ is the only free parameter, which may still depend on system size  $K$. The left inset of Fig.~1 of the main article shows that all the computed data can be mapped on top of each other by rescaling $W\to W/s(A)$. Identifying the values of $a(K)$ independently for different system sizes $K$, this allows us to extrapolate $a(K)$ to larger system sizes that cannot be treated numerically. In particular, by fitting the function $a(K)$ in the right inset of Fig.~1 of the main article, we expect $a(K)$ to saturate at $a_\infty=3.55 \pm 0.25$. However, we note that the small regime of accessible system sizes does not allow us to distinguish between an exponential and a power-law saturation of $a(K)$. Assuming  $a(K)$ does not saturate faster than an exponential function but at the same time not slower than a power-law, this results in the given uncertainty of $a_\infty$. Now, the critical disorder strength, at which the transition occurs, is given by
\begin{equation}
W_c(A) = W_c^\text{Heis}  \sqrt{1 + (A/a_\infty)^2},
\end{equation} 
where $W_c^\text{Heis}\approx 3.7$ is the critical disorder strength of the random field Heisenberg chain.

\begin{figure}
\centering
\includegraphics[width=0.98\linewidth]{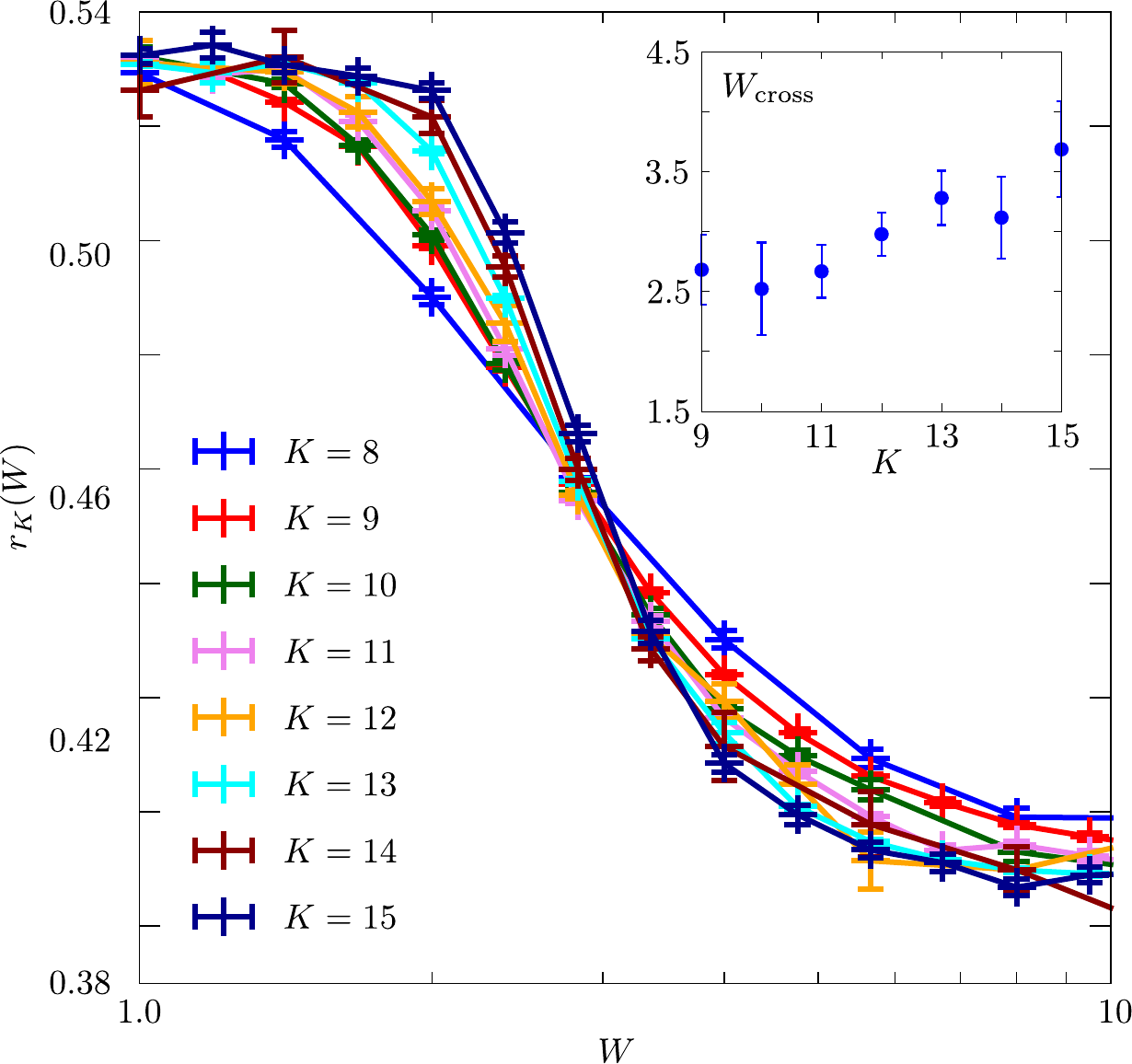}
\caption{Flow of the level statistics. For a fixed value of $A$, the functions $r_K(W)$ are shifted to larger values of disorder strength with increasing system sizes $K$. This is quantified by the disorder $W_\text{cross}$, at which two functions $r_K(W)$ and $r_{K-1}(W)$ cross. The behavior of $W_\text{cross}$ is shown in the inset. Data is generated for $A=1.0$.}
\label{fig:SOM_flow}
\end{figure}

We note that we observe a flow of $r_K(W)$ (for constant $A$) to larger disorder values if the system size is increased, see Fig.~\ref{fig:SOM_flow}. This typical feature of MBL systems can be used to extrapolate the critical disorder strength $W_c$ at infinite system size. However, we could not apply this method in our model due to too large error bars.

\subsection{Non-extensive magnetization exchange and memory about the initial state}

In the main text, we relate the logarithmic contribution $k A^2/W^2\ln(t)$ to the entanglement entropy $S_\mathcal{A}(t)$ to fluctuations of magnetization induced by the central spin. Here, we want to analyze this mechanism in more detail.

\begin{figure}
\centering
\includegraphics[width=\linewidth]{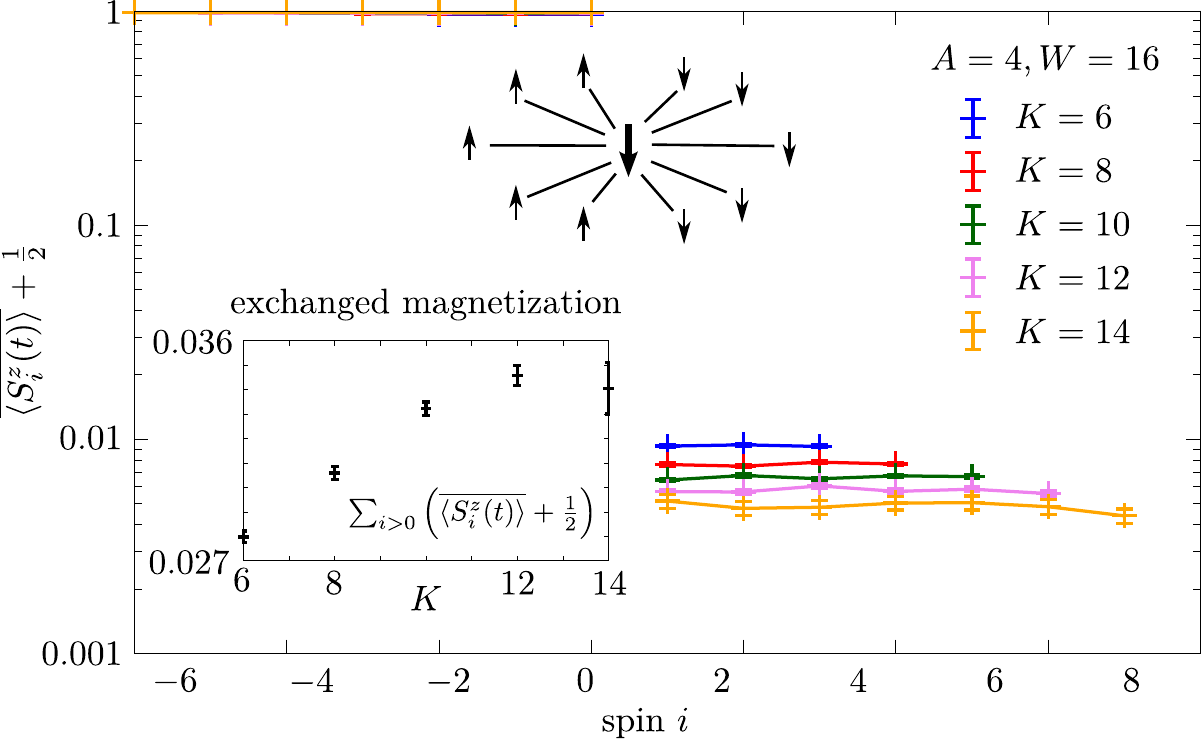}
\caption{Non-extensive exchange of magnetization in a central spin model. For $i>0$, we see the transported magnetization per spin, which decreases with increasing system size $K$. However, the inset shows that the total amount of transferred magnetization will saturate, motivating the term 'non-extensive' transport. For the data shown in this plot, we set $J=0$ in order to study solely magnetization changes that are due to the central spin.}
\label{fig:transport}
\end{figure}

First, we want to emphasize that, by magnetization exchange we do not mean extensive transport, like in a conducting or ergodic phase, which would be in contradiction to MBL. In fact, the observed magnetization exchange is non-extensive, i.e. the total amount of magnetization transferred through the central spin is independent of system size. To illustrate this feature, let us use an initial state where all spins in the left half of the system $\mathcal{L}$ are polarized in the positive $z$ direction, and all spins in the right half $\mathcal{R}$ are polarized in the opposite direction, $\ket{\psi} = \ket{\uparrow \uparrow \uparrow \ldots \downarrow \downarrow \ldots}$. Let us further set $J=0$, such that spins only interact with each other via the central spin, which couples with strength $A/K$. For $A=0$ only the disorder term remains and  $\ket{\psi}$ is an eigenstate of the system. Increasing $A$ to finite values, resonances occur and spins of different halves of the system mix, such that positive magnetization from $\mathcal{L}$ will be transmitted through the central spin (starting in $\ket{\downarrow\,}$) to $\mathcal{R}$, or vice versa. In order to quantify this amount, we make use of the infinite time average 
\begin{equation}
\overline{\mean{I^z_i(t)}} = \lim_{T\to\infty} \frac{1}{T} \int_0^T \tot{t} \matrixel{\psi(t)}{I^z_i}{\psi(t)}
\end{equation}
for each spin $i$. We compute the value of $\overline{\mean{I^z_i(t)}} = \tr{\bar{\rho} I^z_i}$ exploiting the time-averaged density matrix 
\begin{align}
\overline{\rho}&=\lim_{T\to\infty}\frac{1}{T}\int_0^T \tot{t} \ket{\psi(t)}\bra{\psi(t)}\\
&=\sum_{i} (\rho^E)_{ii} \ket{E_i}\bra{E_i}\nonumber.
\end{align}
We illustrate the results of this analysis in Fig.~\ref{fig:transport}. For the left half of the spin chain, we identify memory about the initial state, which accounts to MBL. However, at the same time, magnetization is leaking to the right half of the chain. Following our perturbative arguments of the NCSM \cite{Note1}, we conjecture that the change of magnetization per spin scales as $A^2/(W^2 K)$, and thus, vanishes in an infinite system. However, as the size of the right half of the chain increases too, the total amount of exchanged magnetization 
\begin{equation}
M=\sum_{i\in \mathcal{R}} \left(\overline{\mean{I^z_i(t)}} + \frac{1}{2}\right)
\end{equation}
will saturate as $K$ increases. This is supported by the inset in Fig.~\ref{fig:transport}. We want to stress that this peculiar phenomenon is a result of scaling the coupling to the central spin as $A/K$. For a coupling strength decaying faster with increasing system size, this effect would be vanishing in the thermodynamic limit.  In contrast, for a slower decaying coupling, the analysis of Ref. \cite{Ponte2017} suggests that the system would always become ergodic. 

\begin{figure}
\centering
\includegraphics[width = \linewidth]{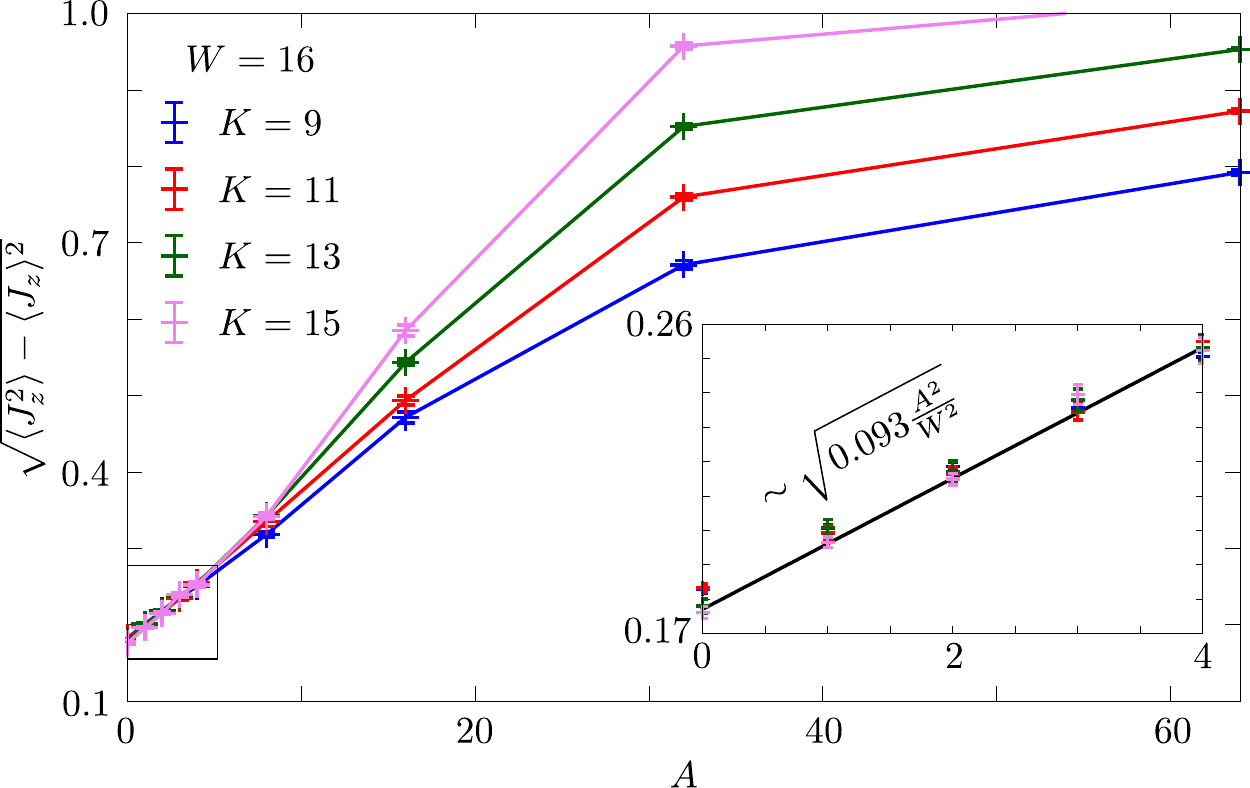}
\caption{Magnetization fluctuation within a bipartition of the central spin model. For $A\lesssim W$ we observe non-extensive behavior, which motivates a localized phase. However, as the fluctuations grow with $A$, we conclude that the coupling to the central spin enables magnetization to move out of the bipartition. The inset magnifies this parameter region and compares the slope with the slope of the logarithmic entanglement growth. For $A\gtrsim W$, we find an extensive magnetization fluctuation, indicating a conducting (ergodic) phase.}
\label{fig:partfluctuations}
\end{figure}

If we now consider the N\'eel state $\ket{\psi}=\ket{\uparrow\downarrow\uparrow\downarrow\ldots}$ as initial state, the non-extensive exchange of magnetization is no longer directed from $\mathcal{L}$ to $\mathcal{R}$ or vice versa. Note that we could still define two sub-lattices $A$ and $B$ in which initially all spins are polarized in the same direction. However, for finite values of $J$, where magnetization may move along the Heisenberg chain within the localization length, one could not separate the effect of the central spin. Therefore, it is more instructive to quantify the exchange of magnetization by means of the magnetization fluctuation
\begin{equation}
\mathcal{F} = \mean{J_z^2} - \mean{J_z}^2,
\end{equation}
where $J_z = \sum_{i\in \mathcal{L}} S_z^i$. As the total magnetization in $z$ direction of the model is conserved, fluctuations within one half of the system indicate motion of magnetization to the other half. In Fig.~\ref{fig:partfluctuations}, we study the fluctuations of eigenstates in the center of the spectrum for different system sizes. For $A=0$, we find a finite value of $\mathcal{F}$ which corresponds to the motion of magnetization at the border between $\mathcal{L}$ and $\mathcal{R}$ within the Heisenberg chain. For $A\gg W$ instead, the system is in an ergodic phase and we observe an extensive increase of fluctuations. Intriguingly, for $0<A\lesssim W$, where the above performed level statistic analysis suggests a localized phase, exchange of magnetization through the central spin is non-extensive, but increases with $A$. In detail, the inset of Fig.~\ref{fig:partfluctuations} shows that the dependence of the magnetization fluctuation on the coupling constant and the disorder strength matches exactly to the slope of the logarithmic growth of entanglement entropy between $\mathcal{L}$ and $\mathcal{R}$ studied in the main article. This indicates that the enhancement of the rate of the entanglement entropy growth is a consequence of an increase of fluctuations of $J_z$ due to the central spin. 

\subsection{Logarithmic entanglement growth}
\begin{figure}
\includegraphics[width=\linewidth]{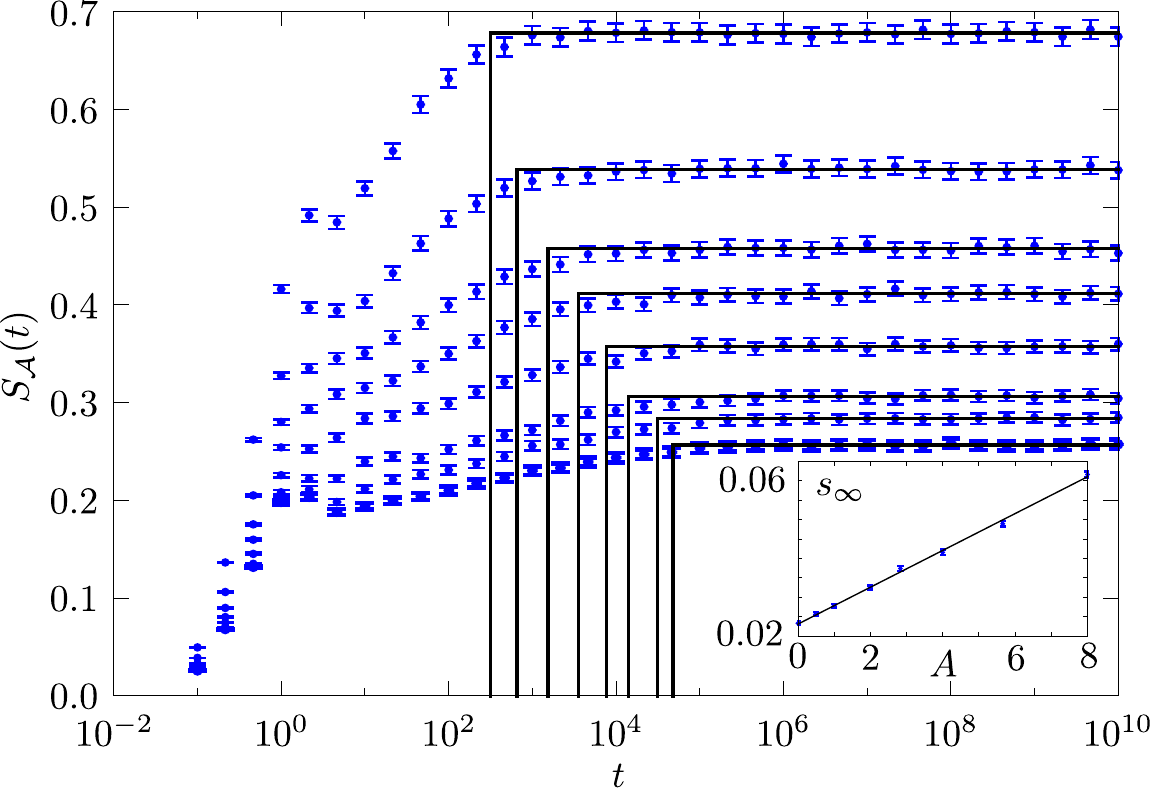}
\caption{Saturation value and saturation times of the entanglement entropy. The inset shows that the saturation value of $S_\mathcal{A}(t)$ increases linearly with  $A$. Saturation times instead reduce. We show data for $K=11, W=16$ and $A\in \{ 0, 0.5, 1.0, 2.0, 2.83, 4.0,5.66,8.0\}$ (bottom to top).}
\label{fig:entsat}
\end{figure}

 While we have addressed the impact on the growth rate of the entanglement entropy in the last section, we address the saturation value and the saturation time of $S_\mathcal{A}(t)$ in this section. 
In Fig.~\ref{fig:entsat}, we show data for $K=11$. Evidently,  the saturation value per spin $\hat{s}_\infty$ grows only linearly with $A$. This should be compared to the quadratic growth of the slope of $S_\mathcal{A}(t)$ that is studied in the main text. Hence, demanding a functional form
\begin{equation}
S_\mathcal{A}(t) \sim \hat{\xi} \hat{s}_\infty \ln(t),
\end{equation}
as is proposed for typical MBL systems \cite{Huse2014, Serbyn2013}, we conclude that also the parameter $\hat{\xi}$ is modified by the coupling to the central spin. 
This behavior can be expected: The value $\hat{s}_\infty$ has by construction no impact on the saturation time $t_S$ of $S_\mathcal{A}(t)$, while Fig.~\ref{fig:entsat} shows that the time scales of $t_S$ are significantly reduced with increasing $A$. 
However, we want to stress that $\hat{\xi}$ should no longer be interpreted as a localization length in our non-local model. In fact, $\hat{\xi}$ quantifies both in the absence and in the presence of the central spin the amount of spins that are able to dephase with a given spin. Thus, we conjecture that the increase in $\tilde{\xi}$ with $A$ is due to the increasing number of coupled spins.

\subsection{Area law of entanglement entropy}

\begin{figure}
\includegraphics[width=\linewidth]{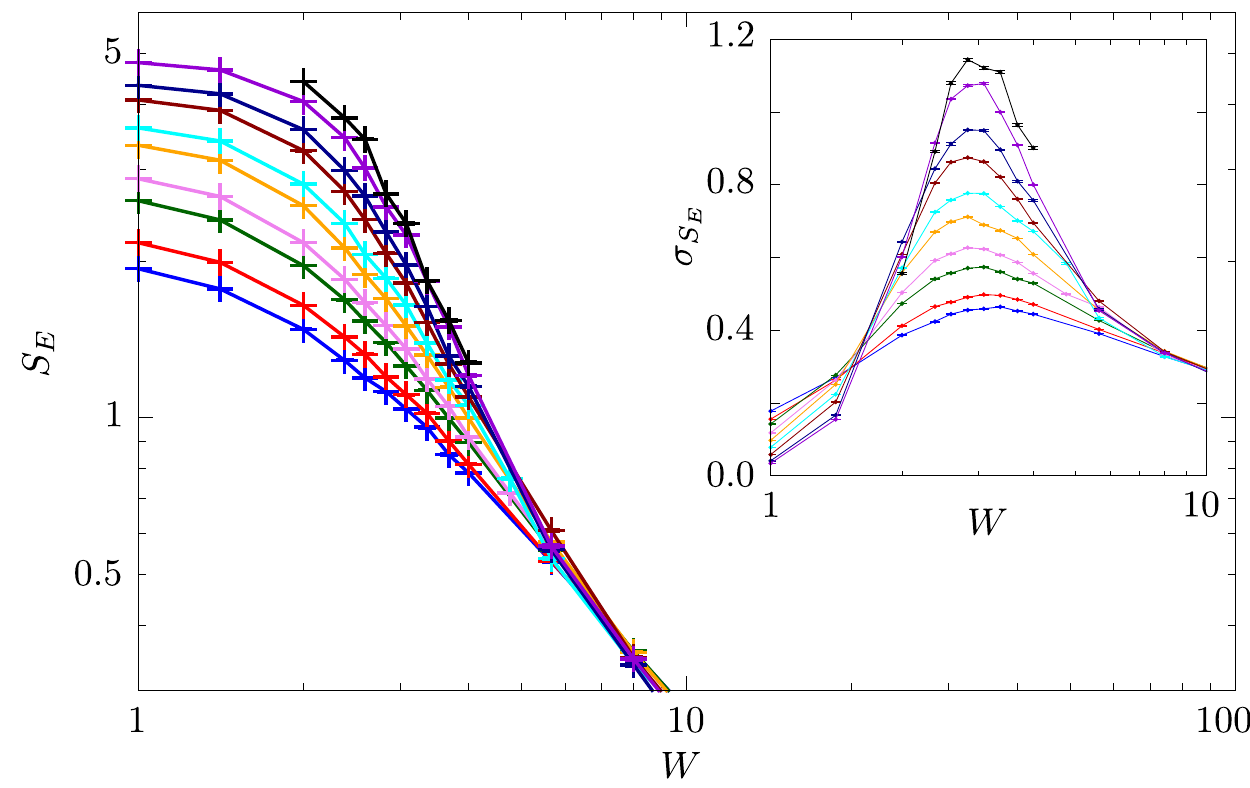}
\caption{Area vs. volume law of entanglement entropy. We show data for $A=1$ and $7\leq K \leq 16$ (bottom to top) in the center of the eigenvalue spectrum. Deep in the localized phase, $W \gg A$, the eigenstates lead to a entanglement entropy $S_E$ that is small and independent of $K$. Instead, $S_E$ grows linearly with $K$ once the system delocalizes at $W \ll A$. Around $W\approx 3$ we observe the MBL transition indicated by a peaking standard deviation of $S_E$ (see inset). There, both localized and delocalized eigenstates contribute, leading to a standard deviation that grows linearly with $K$.}
\label{fig:area}
\end{figure}

Next, we study whether the many-body eigenstates in the sector with zero total spin have volume- or area-law entanglement entropy.
The entanglement entropy $S_E = -\tr{\rho^E_\mathcal{A} \ln \rho^E_\mathcal{A}}$ of an eigenstate $\ket{E}$, where $\rho^E_\mathcal{A}=\tr[\mathcal{B}]{ \ket{E}\bra{E} }$ , quantifies how localized an eigenstate is. While thermalized systems show a volume law, i.e. $S_E\sim L^{d}$, where $d$ is the dimension of the system, localized models obey an area law, $S_E\sim L^{d-1}$. Fig.~\ref{fig:area} illustrates that the central spin model consists of eigenstates that show an area law deep in the localized regime, regardless of the logarithmic transport properties. We conclude that despite the presence of resonances via the central spin that enable transport, these resonances remain rare and unable to delocalize the system. 

\subsection{Power law of the correlation function}
\begin{figure}
\includegraphics[width=\linewidth]{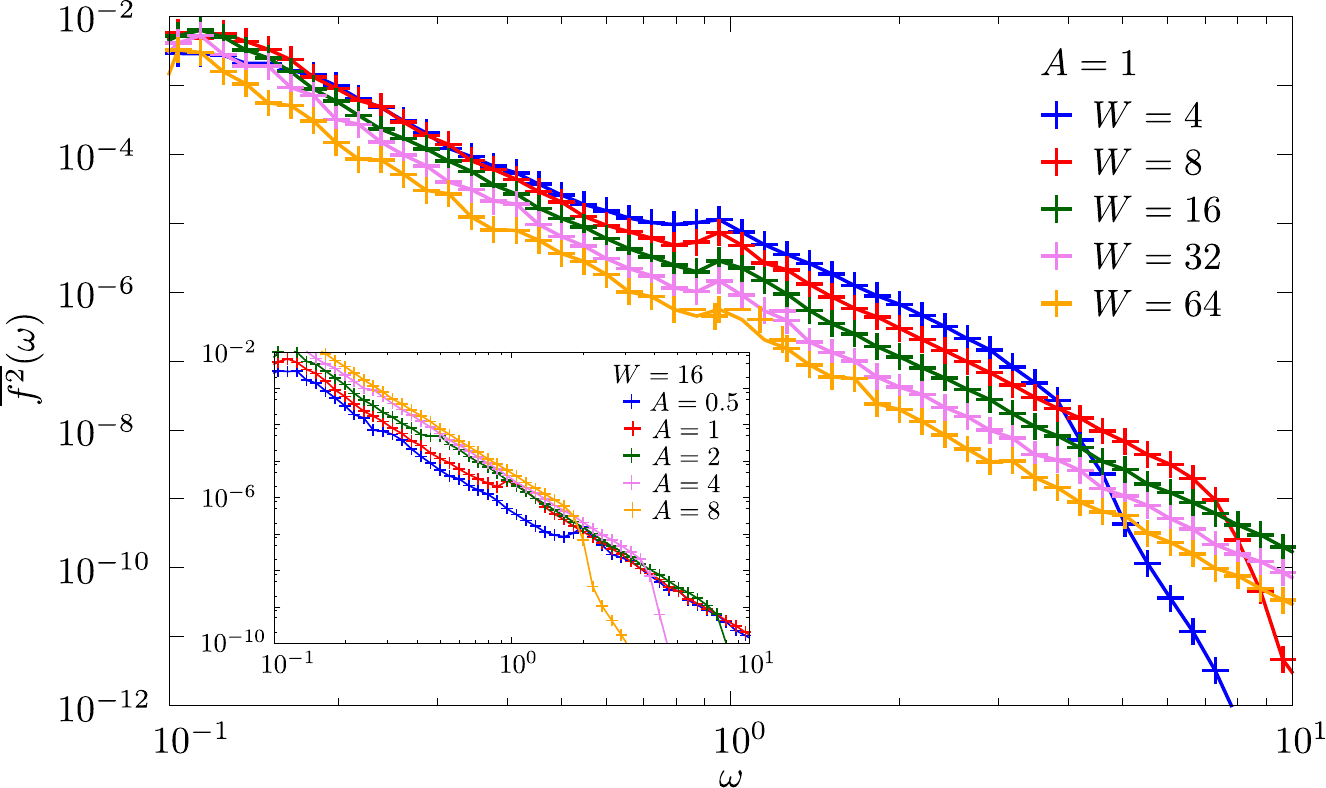}
\caption{Power law decay of $\overline{f^2}(\omega)$ for various values of $A$ and $W$ at $K=9$. Each function decays as $\sim \omega^{-4}$. }
\label{fig:eftpowerlaw}
\end{figure}

In the main text, we discuss the power law behaviors of $f^2(\omega)$, which is the Fourier transformed auto-correlation function of $S_z(t)$. In particular, we have claimed that, for $\omega \gtrsim A/K$ and in the localized phase, $f^2(\omega) \sim \omega^{-4}$ holds  independently of the disorder strength $W$ and coupling strength $A$. In Fig.~\ref{fig:eftpowerlaw}, we support this statement with numerical data over a wide range of parameters.
 However, we note that $f^2(\omega)$ is expected to become non-self averaging in the MBL phase~\cite{Serbyn2017}. Hence, the value of exponent may depend on the averaging procedure.

\end{document}